\documentclass[%
 aip,
rsi,%
 amsmath,amssymb,
 reprint,%
longbibliography,
]{revtex4-1}

\usepackage{xcolor, soul}
\sethlcolor{yellow}
\usepackage{graphicx}% Include figure files
\usepackage{dcolumn}% Align table columns on decimal point
\usepackage{bm}% bold math
\usepackage{euscript}
\usepackage{tikz}
\newcommand*\circled[1]{\tikz[baseline=(char.base)]{
    \node[shape=circle, draw, inner sep=1pt, 
        minimum height=10pt] (char) {#1};}}
        
\begin{document}

\preprint{AIP/123-QED}

\title[Magnetically induced topological transitions]{Magnetically induced topological transitions of hyperbolic dispersion in biaxial gyrotropic media}
% Force line breaks with \\
%\thanks{Footnote to title of article.}
\author{Vladimir~R.~Tuz}
 \email{tvr@jlu.edu.cn}
\affiliation{State Key Laboratory of Integrated Optoelectronics, College of Electronic Science and Engineering, International Center of Future Science, Jilin University, 2699 Qianjin Street, Changchun 130012, China}
\author{Volodymyr~I.~Fesenko}
 \email{volodymyr.i.fesenko@gmail.com}
\affiliation{State Key Laboratory of Integrated Optoelectronics, College of Electronic Science and Engineering, International Center of Future Science, Jilin University, 2699 Qianjin Street, Changchun 130012, China}
\affiliation{Institute of Radio Astronomy of National Academy of Sciences of Ukraine, 4 Mystetstv Street, Kharkiv 61002, Ukraine} 

\date{\today}% It is always \today, today,
             % but any date may be explicitly specified

\begin{abstract}
Magnetically induced topological transitions of isofrequency surfaces of bulk waves propagating through an unbounded biaxial gyrotropic medium are studied. The medium is constructed from a two-component superlattice composed of magnetized ferrite and semiconductor layers. To derive the constitutive parameters of the gyrotropic medium, a homogenization procedure from the effective medium theory is applied. The study is carried out in the frequency range near the frequency of ferromagnetic resonance, where the magnetic subsystem possesses the properties of natural hyperbolic dispersion. The topological transitions from an open type-I hyperboloid to several intricate hyperbolic-like forms are demonstrated for the extraordinary waves. We reveal how realistic material losses change the form of isofrequency surface characterizing hyperbolic dispersion. The obtained results broaden our knowledge on the possible topologies of isofrequency surfaces that can appear in gyrotropic media influenced by an external static magnetic field.
\end{abstract}

%Valid PACS numbers may be entered using the \verb+\pacs{#1}+ command.

%\pacs{Valid PACS appear here}% PACS, the Physics and Astronomy Classification Scheme.

%\keywords{Chiral media; Nanomaterials; Plasmonics; Optical activity; Dichroism; Circular polarizers}
%Use showkeys class option if keyword display desired

\maketitle

\section{Introduction}
\label{sec:intro}

Hyperbolic metamaterials can serve as a novel functional platform for waveguiding, imaging, sensing, quantum and thermal engineering beyond conventional devices. These metamaterials derive their name from the hyperbolic topology of isofrequency surfaces characterizing dispersion features of the medium. Hyperbolic metamaterials utilize the concept of engineering the basic dispersion relations of waves to provide unique electromagnetic modes that can have a broad range of applications (see a comprehensive review on the theory and application of hyperbolic metamaterials in Refs.~\onlinecite{Poddubny_NatPhoton_2013, Jacob_NanoConv_2014, Ferrari_Prog_2015, Davidovich_UFN_2019, Takayama_JOptSocAmB_2019, Guo_JApllPhys_2020}). 

A hyperbolic metamaterial is associated with an extremely anisotropic medium in which the principal components of permittivity or permeability tensor have different signs. In the ideal case, in the absence of dissipation and spatial dispersion, the isofrequency surface of the Fresnel's equation for such a medium acquires an open form of the one-fold or two-fold hyperboloid of revolution.\cite{Smith_PhysRevLett_2003} This form of the dispersion topology differs drastically from the usual case of closed elliptical dispersion typical for conventional anisotropic media.\cite{landau_1960_8} Metallic wire composites\cite{Simovski_AdvMat_2012} and planar metal-dielectric finely-stratified structures (superlattices)\cite{Schilling_PhysRevE_2006, Zhukovsky_OptExpress_2013} are two common designs of metamaterials demonstrating hyperbolic dispersion. The optical properties of such artificial structures can be described in the framework of the effective medium theory (EMT).\cite{Zhukovsky_PhysRevA_2012}

The most important property of hyperbolic metamaterials is related to the behavior of waves with large magnitude wavevectors (high-$k$ waves). In vacuum, such high-$k$ waves are evanescent modes which decay exponentially. However, in hyperbolic media the open form of the isofrequency surface allows for propagating waves with large wavevectors.\cite{Jacob_OptExpress_2006} The possibility of propagating high-$k$ waves leads to several unusual effects, including strong enhancement of spontaneous emission,\cite{Poddubny_PhysRevB_2013, Sreekanth_JApplPhys_2013} broadband infinite density of states,\cite{Jacob_APL_2012} super-resolution imaging,\cite{Liu_Science_2007, Novitsky_AnnPhys2017} and abnormal scattering.\cite{Shen_PhysRevX_2015}

The ability to tune and switch the properties of hyperbolic metamaterials can greatly broaden their application in many fields.\cite{Smalley_AdvOptPhoton_2018} In particular, in hyperbolic metamaterials, it is important to gain control over the topological transitions, which may allow performing switching between hyperbolic and elliptic dispersion states. The most straightforward way to realize such a control is to use magneto-active components in the metamaterial composition and an external static magnetic field as a driving agent.\cite{Li_ApplPhysLett_2012, Macedo_2017, Kolmychek_OptLett_2018, Pomozov_PhysSolidState_2018, Fedorin_Super_2018, Yan_Plasmon_2019}

However, the application of an external static magnetic field modifies the property of the medium, which becomes gyrotropic. A gyrotropic medium is characterized by permittivity or permeability tensor having antisymmetric off-diagonal parts. Gyrotropy changes the dispersion characteristics of the medium and causes the nonlocality\cite{Zhang_OptExpress_2019} and nonreciprocity effects.\cite{Leviyev_APLPhoton_2017} Moreover, magneto-optic materials can exhibit natural hyperbolic dispersion. In particular, in gyromagnetic materials (e.g., ferrites), the hyperbolic dispersion originates from the ferromagnetic resonance,\cite{Lan_ApplPhysLett_2015, Lokk2017} whereas in gyroelectric materials (e.g., semiconductors), it appears due to the plasma resonance.\cite{akhiezer1975plasma} Therefore, the simultaneous presence of natural and engineered hyperbolic dispersion in a complex structure can lead to the appearance of very intricate forms of isofrequency surfaces\cite{Tuz_OptLett_2017, Fesenko_PhysRevB_2019} (see Refs.~\onlinecite{Depine_JOSAA_2006, Durach_ApplScien_2020} for a complete taxonomy of isofrequency surfaces that can be realized in uniaxial anisotropic and bianisotropic optical materials, respectively).

It is obvious that combining gyroelectric and gyromagnetic materials into a single gyroelectromagnetic system can bring many unique dispersion features, which are unattainable in separate subsystems.\cite{Kaganov_PhysUsp_1997, Tarkhanyan_PSSb_2008, Tarkhanyan_JMMM_2010, Tuz_Springer_2016, Fesenko_OptLett_2016, Tuz_JApplPhys_2017, Tuz_Superlattice_2017, Farhadi_ApplPhysA_2019} In particular, in the present paper, we demonstrate that in a biaxial gyrotropic medium composed of magnetized ferrite and semiconductor layers, some specific distortions of isofrequency surfaces may occur. These distortions manifest themselves near the frequency of ferromagnetic resonance, where the magnetic subsystem possesses the properties of natural hyperbolic dispersion. We first study the idealized lossless structure and then analyze the effect of actual losses on the hyperbolic dispersion. Thus, we show that in composite structures containing magneto-active components, in addition to the possibility of control, one can obtain a diversity of open hyperbolic-like topologies of isofrequency surfaces.

\section{Dispersion equation}
\label{sec:theory}

In what follows, we study characteristics of plane electromagnetic waves (bulk waves \cite{Nayfeh_bulk_1995}) propagating in an arbitrary direction through an infinite homogeneous gyrotropic medium. The gyrotropic medium is magnetized up to the saturation level in the presence of a uniform magnetic field $\vec M$. We introduce the Cartesian coordinate system with the $z$ axis directed along the vector $\vec M$. In the chosen framework, the magnetized gyrotropic medium can be characterized by the second-rank tensors of permittivity and permeability 
\begin{equation}
\hat \varepsilon=\left( {\begin{matrix}
   {\varepsilon_{xx}} & {\varepsilon_{xy}} & 0 \cr
   {-\varepsilon_{xy} } & {\varepsilon_{yy} } & 0 \cr
   0 & 0 & {\varepsilon_{zz}} \cr
\end{matrix}
} \right),~
\hat\mu=\left( {\begin{matrix}
   {\mu_{xx} } & {\mu_{xy}} & 0  \cr
   {-\mu_{xy} } & {\mu_{yy}} & 0  \cr
   0 & 0 & {\mu_{zz} }  \cr
 \end{matrix}
} \right), \label{eq:eff}
\end{equation}
which establish the relations between the electric and magnetic fields and inductions as
\begin{equation}
 \left( {\begin{matrix}
   {\vec{D}} \cr
   {\vec{B}} \cr
\end{matrix}
} \right) =  \left( {\begin{matrix}
   {\hat\varepsilon} & {0} \cr
   {0} & {\hat\mu} \cr
\end{matrix}
} \right) \left( {\begin{matrix}
   {\vec{E}} \cr
   {\vec{H}} \cr
\end{matrix}
} \right).
\label{eq:const}
\end{equation}

Time and space harmonic variations of the electric ($\vec E$) and magnetic ($\vec H$) components of the wave are given by
\begin{equation}
\vec E (\vec H) = \vec E_0 (\vec H_0)\exp\left[i(-\omega t+k_xx+k_yy+k_zz)\right],
\label{eq:EandH}
\end{equation}
where $k_x$, $k_y$, and $k_z$ are projections of the wavevector $\vec k$ in the Cartesian coordinates, and $\omega$ is the angular frequency.

Starting from a pair of the curl Maxwell’s equations $\nabla \times \vec E = i k_0\vec B$ and $\nabla\times\vec H = -i k_0\vec D$ in the absence of sources in the volume of medium ($\nabla \cdot \vec D = 0$ and $\nabla \cdot \vec B = 0$), one can derive a system of two coupled wave equations for the $z$-components of the electromagnetic field:\cite{Gurevich_book_1963}
\begin{align}
 &\left( {\begin{matrix}
   {\xi\zeta+\zeta k_z^2\varepsilon_{zz}-k_0^2\xi\varepsilon_{zz}\mu_\bot} & {k_0 k_z\mu_{zz}\left(\chi-k_x k_y\eta \right)} \cr
   {-k_0 k_z\varepsilon_{zz}\left(\chi-k_x k_y\eta \right)} & {\xi\zeta+\xi k_z^2\mu_{zz}-k_0^2\zeta\mu_{zz}\varepsilon_\bot} \cr
\end{matrix}
} \right) \nonumber \\
&\qquad\qquad\qquad\qquad\qquad\qquad\qquad\times\left( {\begin{matrix}
   {E_z} \cr
   {H_z} \cr
\end{matrix}
} \right) = 0,
\label{eq:wave_eq}
\end{align}
where $k_0=\omega/c$, $c$ is the speed of light in vacuum, $\eta=\varepsilon_{xx}\mu_{yy}-\varepsilon_{yy}\mu_{xx}$, $\xi=k_x^2\varepsilon_{xx}+k_y^2\varepsilon_{yy}$, $\zeta=k_x^2\mu_{xx}+k_y^2\mu_{yy}$, $\chi=\zeta\varepsilon_{xy}+\xi\mu_{xy}$, and $\varepsilon_{\bot} = \varepsilon_{xx}\varepsilon_{yy} + \varepsilon_{xy}^2$ and $\mu_{\bot} = \mu_{xx}\mu_{yy} + \mu_{xy}^2$ are two generalized transverse effective constitutive parameters of the gyrotropic medium.

The nontrivial solution of system (\ref{eq:wave_eq}) exists when its determinant of coefficients equals to zero. Under this condition, the biquadratic equation describing propagation of bulk waves through the gyrotropic medium is obtained in the form:\cite{Tuz_OptLett_2017, Fesenko_PhysRevB_2019}
\begin{align}
&(\varepsilon_{zz}\mu_{zz})^{-1}\bigl\{k_x^4\varepsilon_{xx}\mu_{xx}+k_y^4\varepsilon_{yy}\mu_{yy}+k_z^4\varepsilon_{zz}\mu_{zz}+k_x^2k_y^2 \nonumber \\
&\times(\varepsilon_{xx}\mu_{yy}+\varepsilon_{yy}\mu_{xx})+k_x^2k_z^2(\varepsilon_{xx}\mu_{zz}+\varepsilon_{zz}\mu_{xx})+k_y^2k_z^2 \nonumber \\
&\times(\varepsilon_{yy}\mu_{zz}+\varepsilon_{zz}\mu_{yy})- 
k_0^2\bigl[ k_x^2(\varepsilon_{xx}\varepsilon_{zz}\mu_{\bot}+\mu_{xx}\mu_{zz}\varepsilon_{\bot}) \nonumber \\
&+k_y^2(\varepsilon_{yy}\varepsilon_{zz}\mu_{\bot}+\mu_{yy}\mu_{zz}\varepsilon_{\bot})+k_z^2\varepsilon_{zz}\mu_{zz}(\varepsilon_{xx}\mu_{yy} \nonumber \\
&+\varepsilon_{yy}\mu_{xx}-2\varepsilon_{xy}\mu_{xy})\bigr] \bigr\}+k_0^4\varepsilon_{\bot}\mu_{\bot}=0.
\label{eq:disp_eq}
\end{align} 
This equation is also known as the Fresnel's equation for wave normals.\cite{landau_1960_8} It implicitly determines the dispersion relation, i.e., the frequency as a function of the wavevector. At a fixed frequency $\omega$, it defines the wavevector surface (isofrequency contour) given in the $k$-space.

Using the spherical coordinate system in which $k_x=k\sin\theta\cos\varphi$, $k_y=k\sin\theta\sin\varphi$, and $k_z=k\cos\theta$, Eq. (\ref{eq:disp_eq}) can be compactly rewritten in therms of $\kappa=k/k_0$ as
\begin{equation}
\EuScript A\kappa^4 + \EuScript B\kappa^2 + \EuScript C = 0,
\label{eq:disp_eq_k}
\end{equation}
where $\EuScript A=(\varepsilon_{zz}\mu_{zz})^{-1} (\overline{\varepsilon} \sin^2\theta + \varepsilon_{zz} \cos^2\theta) (\overline{\mu} \sin^2\theta + \mu_{zz} \cos^2\theta)$, $\EuScript B=-[(\varepsilon_{xx}\mu_{yy} + \mu_{xx}\varepsilon_{yy} - 2\varepsilon_{xy}\mu_{xy})\cos^2\theta +(\varepsilon_{zz}\mu_{zz})^{-1}(\varepsilon_{\bot}\overline{\mu}\mu_{zz}+\mu_{\bot}\overline{\varepsilon}\varepsilon_{zz})\sin^2\theta]$, $\EuScript C=\varepsilon_{\bot}\mu_{\bot}$, $\overline{\varepsilon} = \varepsilon_{xx}\cos^2\varphi + \varepsilon_{yy}\sin^2\varphi$, and  $\overline{\mu} = \mu_{xx}\cos^2\varphi + \mu_{yy}\sin^2\varphi$. 

Solution of Eq.~(\ref{eq:disp_eq_k}) is straightforward:
\begin{equation}
\kappa = \pm\sqrt{\frac{\EuScript B \pm \sqrt{\EuScript B^2-4\EuScript {AC}}}{2\EuScript A}}.
\label{eq:solution_eq_k}
\end{equation}

When the dispersion characteristics of the components of tensors (\ref{eq:eff}) are known, solution (\ref{eq:solution_eq_k}) can be used to calculate the isofrequency surfaces for the bulk waves propagating through the infinite homogeneous gyrotropic medium.

\section{Homogenization conditions}
\label{sec:homo}

\begin{figure}
\centerline{\includegraphics[width=0.9\linewidth]{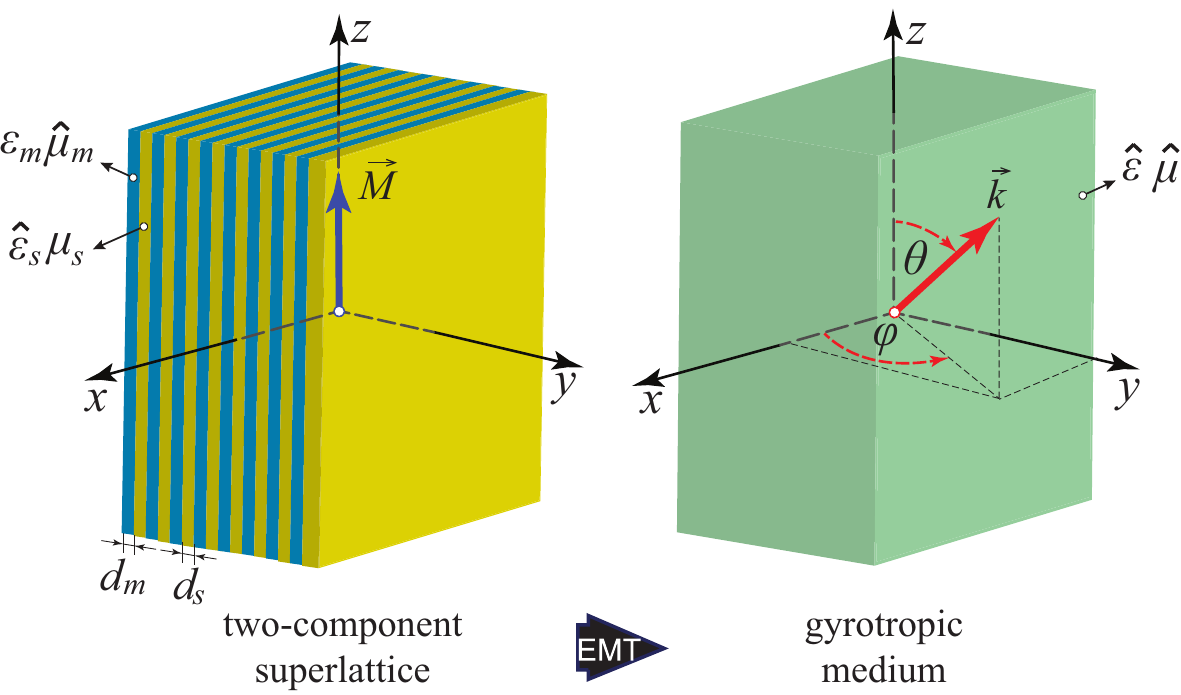}}
\caption{The problem sketch related to both a two-component superlattice influenced by an external static magnetic field $\vec{M}$ and resulting homogenized gyrotropic medium.}
\label{fig:fig1}
\end{figure}

In order to realize a gyrotropic medium, we consider a realistic finely-stratified structure made in the form of a two-component superlattice. In particular, such a structure can be realized for operating in the microwave part of spectrum. \cite{Schuller_JMMM_1999, Chen_JMMM_2006, Lyubchanskii_JApplPhys_2014} It is composed of ferrite (with constitutive parameters $\varepsilon_m$, $\hat \mu_m$) and semiconductor (with constitutive parameters $\hat \varepsilon_s$, $\mu_s$) layers with thicknesses $d_m$ and $d_s$, respectively. The layers are periodically arranged along the $y$ axis. The structure's period is $L = d_m + d_s$, and the number of periods in the superlattice is large enough that it can be considered infinite. We suppose that all dimensions $d_m$, $d_s$, and $L$ are much smaller than the wavelength in the corresponding layer $d_m\ll \lambda$, $d_s \ll \lambda$, and period $L \ll \lambda$ of the superlattice (the long-wavelength limit). Both ferrite and semiconductor subsystems are magnetized uniformly by an external static magnetic field $\vec M$ directed along the $z$ axis transversely to the structure periodicity, as shown in Fig.~\ref{fig:fig1}.

Taking into account the long-wavelength limit, the finely-stratified structure can be equivalently represented by a homogeneous anisotropic medium.\cite{rytov1956} In such a quasi-static approximation, the homogenization procedure \cite{Agranovich_SolidStateCommun_1991} from the effective medium theory is applied to derive components of tensors (\ref{eq:eff}) in an explicit form. They can be obtained by substituting $\mu \to \EuScript G$ and $\varepsilon \to \EuScript G$ to the tensor\cite{Tuz_JMMM_2016, Fesenko_book_2018} 
\begin{equation}
\EuScript {\hat G}=\left( {\begin{matrix} {\EuScript G_{xx}} & {\EuScript G_{xy}} & {0} \cr {-\EuScript G_{xy}} &
{\EuScript G_{yy}} & {0} \cr {0} &
{0} & {\EuScript G_{zz}} \cr
\end{matrix}
} \right)
\label{eq:gem}
\end{equation}
with components $\EuScript G_{xx} = g^{(m)}_{xx}\delta_m+g^{(s)}_{xx}\delta_s+ (g^{(m)}_{xy}-g^{(s)}_{xy})^2\delta_m\delta_s\EuScript D$, $\EuScript G_{xy} = (g^{(m)}_{xy}g^{(s)}_{yy}\delta_m + g^{(s)}_{xy}g^{(m)}_{yy}\delta_s)\EuScript D$, $\EuScript G_{yy} = g^{(m)}_{yy}g^{(s)}_{yy}\EuScript D$, $\EuScript G_{zz}=g^{(m)}_{zz}\delta_m+g^{(s)}_{zz}\delta_s$,  $\delta_m=d_m/L$, $\delta_s=d_s/L$, $\delta_m+\delta_s=1$, $\EuScript D = (g^{(s)}_{yy}\delta_f+g^{(m)}_{yy}\delta_s)^{-1}$, where the expressions for parameters of the tensors $\hat g^{(j)}$ are given in Refs.~\onlinecite{Gurevich_book_1963, Collin_book_1992, akhiezer1975plasma}, and Appendix~\ref{sec:append}. These expressions take into account the resonant nature of dispersion characteristics of the constitutive parameters of ferrite and semiconductor subsystems caused by the applied external static magnetic field. To verify the applicability of our approximation, the results of the homogenization procedure are checked against those of the rigorous transfer-matrix technique,\cite{Tuz_PIERB_2012, Tuz_JOpt_2015} for which a good correlation is found.

For the homogenized medium, all principal components of tensors (\ref{eq:eff}) are different ($\EuScript G_{xx}\ne \EuScript G_{yy}\ne \EuScript G_{zz}$), and the off-diagonal component is a nonzero value ($\EuScript G_{xy} \ne 0$). This corresponds to the conditions of a biaxial gyrotropic crystal.\cite{Mackay_book} Two axes of anisotropy are conditioned by the simultaneous effect of both structure periodicity and external static magnetic field influence. The presence of two axes of anisotropy supplemented by gyrotropy leads to several peculiarities in the dispersion characteristics of the medium. Especially, these peculiarities are most pronounced in the frequency range near the frequency of ferromagnetic resonance. In this range the principal components of the permeability tensor vary from positive to negative values which results in the distinctive topological transitions of the isofrequency surfaces, which are the subject of our following numerical investigation. 

\section{Topological transitions}
\subsection{Lossless case}
\label{sec:lossless}

\begin{figure*}[t]
\centerline{\includegraphics[width=0.9\linewidth]{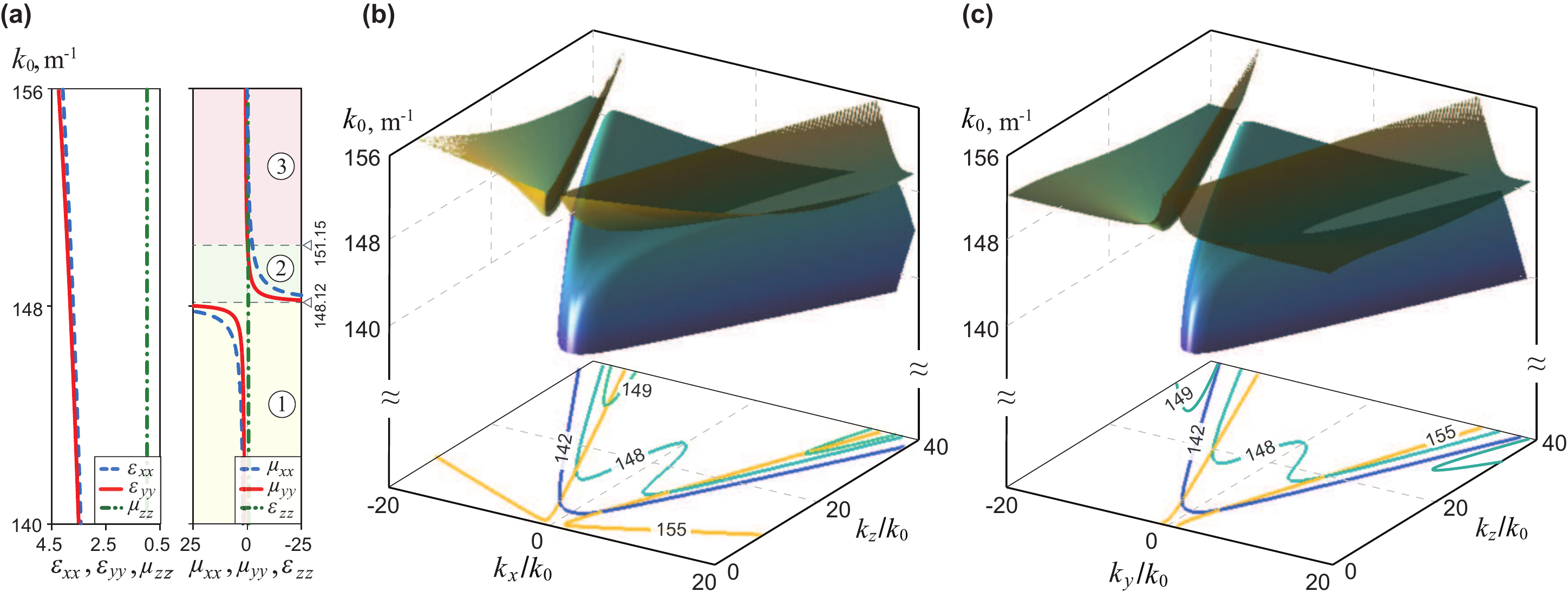}} 
\caption{(a) Dispersion features of principal components of tensors (\ref{eq:eff}) characterizing gyrotropic medium obtained from homogenization of two-component superlattice with parameters $\delta_m = 0.06$ and $\delta_s = 0.94$, and corresponding 3D dispersion relationships and 2D isofrequency contours plotted in (b) $k_x-k_z$ and (c) $k_y-k_z$ planes. All material parameters of the superlattice are derived from Eq.~(\ref{eq:gfs}) for the saturation magnetization of $2930$~G. They are: $\omega_0/2\pi=4.2$~GHz, $\omega_m/2\pi=8.2$~GHz, $\varepsilon_m=5.5$ (for ferrite layers), and $\omega_p/2\pi=10.5$~GHz, $\omega_c/2\pi=9.5$~GHz, $\varepsilon_l=1.0$, $\mu_s=1.0$ (for semiconductor layers).}
\label{fig:fig2}
\end{figure*}

When the direction of the wavevector $\vec k$ is given, Eq.~(\ref{eq:disp_eq_k}) is a quadratic equation for $\kappa^2$. In general, Eq.~(\ref{eq:disp_eq_k}) has two different roots. For a lossless medium, each root takes either a purely imaginary or a purely real value. The real roots correspond to propagating waves. Therefore, two bulk waves with different wavevectors can be propagated in any direction in the medium.\cite{landau_1960_8} Under the accepted notations,\cite{felsen1994radiation} the root in solution (\ref{eq:solution_eq_k}) with upper sign `$+$' is attributed to the `ordinary' waves, while another root with lower sign `$-$' is related to the `extraordinary' waves. In our subsequent consideration, we are mainly interested in the propagation features of extraordinary waves since only these waves demonstrate unusual topological transitions of their isofrequency surface in the chosen frequency range.

The topological form of isofrequency surface depends on the signs of the principal components of tensors (\ref{eq:eff}) characterizing the medium. For the biaxial gyrotropic medium under study, the principal components $\varepsilon_{xx}$ and $\varepsilon_{yy}$ are positive quantities and $\mu_{zz} = 1$ in the entire chosen frequency range. Thus, the dispersion topology is mostly conditioned by values and signs of the remaining three principal components of tensors (\ref{eq:eff}). Among these components, $\varepsilon_{zz}$ is always negative quantity while $\mu_{xx}$ and $\mu_{yy}$ demonstrate resonant behaviors by changing sign. The dispersion features of all principal components of tensors (\ref{eq:eff}) are presented in Fig.~\ref{fig:fig2}(a). In this figure, we distinguish regions of different combinations of positive and negative values of $\mu_{xx}$ and $\mu_{yy}$ by different colors and symbols. They are: Region~ \circled{\small{1}} where $(\mu_{xx}>0) \land (\mu_{yy}>0)$, Region~\circled{\small{2}} where $(\mu_{xx}<0) \land (\mu_{yy}<0)$, and Region~\circled{\small{3}} where $(\mu_{xx}<0) \land (\mu_{yy}>0)$. In addition, the transition points on the $k_0$ scale where components $\mu_{xx}$ and $\mu_{yy}$ change sign are marked by arrows. The point at $k_0\approx 148.12$~m$^{-1}$ corresponds to the frequency of ferromagnetic resonance in the ferrite subsystem.

\begin{figure*}[t!]
\centerline{\includegraphics[width=0.75\linewidth]{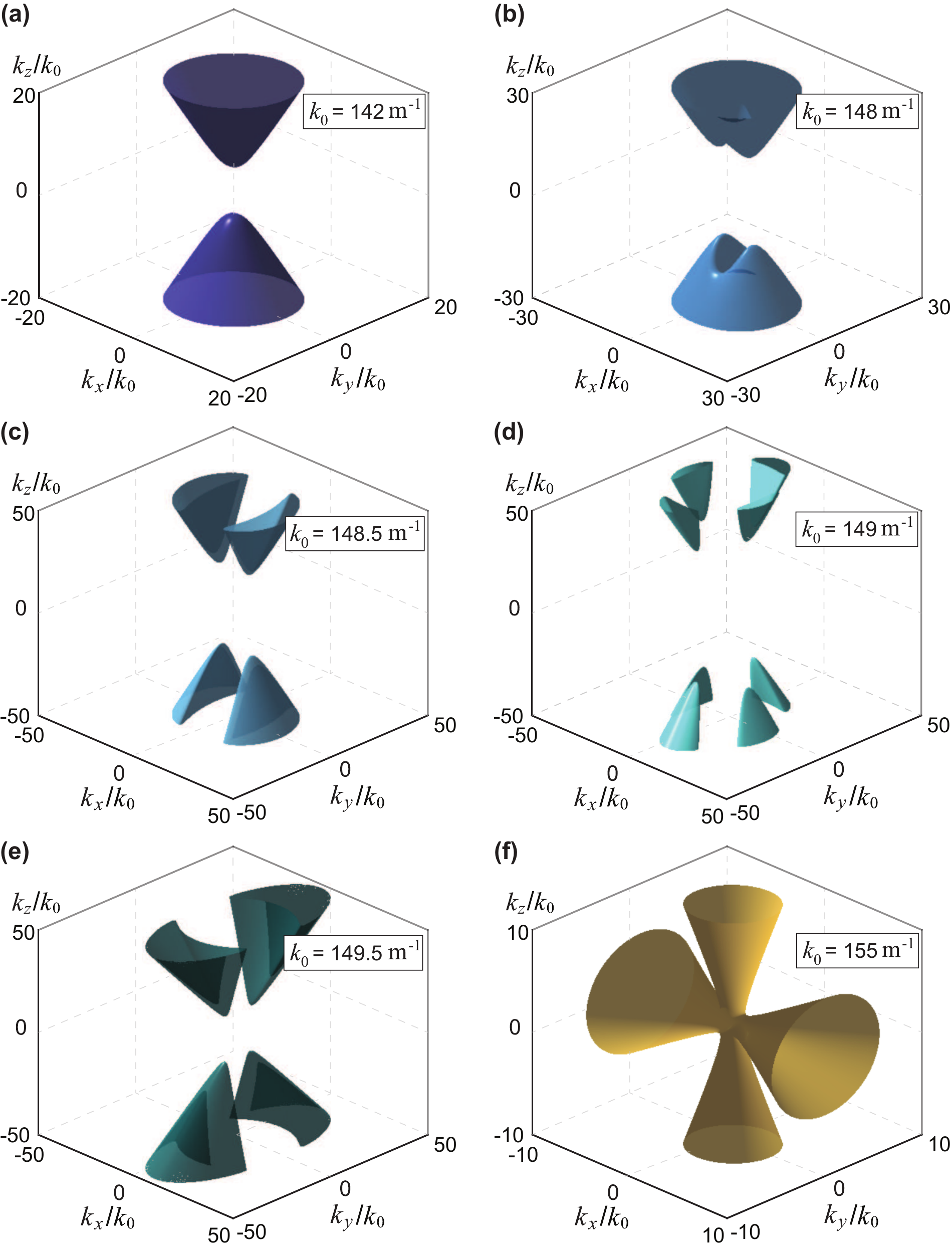}} 
\caption{Topological forms of isofrequency surface related to extraordinary bulk waves propagating through an idealized (lossless) biaxial gyrotropic medium. They correspond to (a), (b) Region~\textcircled{1}, (c)-(e) Region~\textcircled{2}, and (f) Region~\textcircled{3}. All parameters of the superlattice are the same as in Fig.~\ref{fig:fig2}.}
\label{fig:fig3}
\end{figure*}

The corresponding three-dimensional (3D) dispersion relationships of the extraordinary bulk waves propagating through the biaxial gyrotropic medium are plotted in Figs.~\ref{fig:fig2}(b) and \ref{fig:fig2}(c). They are supplemented by several two-dimensional (2D) isofrequency contours, which are drawn at the bottom of these plots. One can see that the topological form of wavevector surface changes drastically during the transition from one region to another. In each region the isofrequency contours are different for the $k_x-k_z$ and $k_y-k_z$ planes since the medium has two axes of anisotropy. Nevertheless, in the entire frequency range, the wavevector surface appears in an open hyperbolic form imposed by the extremely anisotropic tensor $\hat \varepsilon$ having a single negative component ($\varepsilon_{zz}$).

All representative topological forms that arise in the chosen frequency range are collected in Fig.~\ref{fig:fig3}. In particular, in Region~\circled{\small{1}} the isofrequency surface appears as a two-fold (Type I) hyperboloid oriented along the $z$ axis [Fig.~\ref{fig:fig3}(a)]. This form of topology is well known and typical for hyperbolic metamaterials.\cite{Smith_PhysRevLett_2003} When approaching the frequency of ferromagnetic resonance, the isofrequency surface becomes a slightly deformed two-fold hyperboloid [Fig.~\ref{fig:fig3}(b)]. Such a distortion was previously observed for chiral\cite{Gao_PhysRevLett_2015} and gyromagnetic metamaterials,\cite{Chern_OptExpress_2017} although in our case the difference is that the axial symmetry of the hyperboloid is violated by magnetization. When passing through the frequency of ferromagnetic resonance, the surface distortion increases. Specifically, in Region~\circled{\small{2}} the isofrequency surface transits to the form of cones cut into either two or four parts which are oriented along the $z$ axis, as shown in Figs.~\ref{fig:fig3}(c)-\ref{fig:fig3}(e). These changes are caused by the extremely anisotropic tensors $\hat \varepsilon$ and $\hat \mu$, where either one or two components become negative. To the best of our knowledge, such forms of isofrequency surface have not previously been encountered in the literature. Finally, in Region~\circled{\small{3}}, the isofrequency surface appears in the form of two one-fold (Type II) hyperboloids with orthogonal revolution axes [Fig.~\ref{fig:fig3}(f)]. This form of isofrequency surface is attributed to the bi-hyperbolic topology reported recently.\cite{Tuz_OptLett_2017, Fesenko_PhysRevB_2019}  

\begin{figure*}[ht]
\centerline{\includegraphics[width=0.9\linewidth]{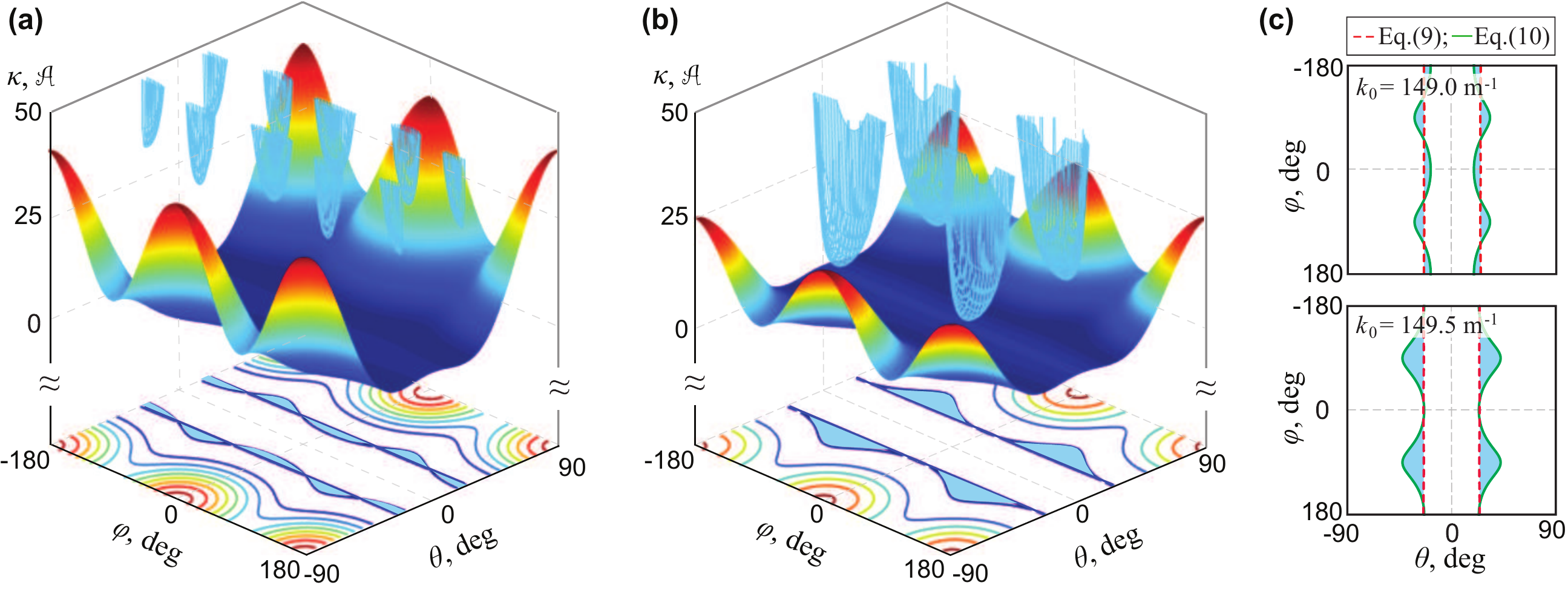}} \caption{Dependence of values of coefficient $\EuScript A$ (color maps) and root $\kappa$ (blue plots) of dispersion equation (\ref{eq:disp_eq_k}) on the propagation direction of extraordinary waves when the frequency parameter is fixed at (a)~$k_0=149.0$~m$^{-1}$ and (b)~$k_0=149.5$~m$^{-1}$. Contours at the bottom are just projections given for an illustrative purpose. (c) A set of solutions of Eqs. (\ref{eq:cond_ep}) and (\ref{eq:cond_mu}) which outlines the corresponding continua of extraordinary waves (blue colored regions) projected on the $\theta-\varphi$ plane. All parameters of the superlattice are the same as in Fig.~\ref{fig:fig2}.}
\label{fig:fig4}
\end{figure*}

Analysis of parameters included in solution (\ref{eq:solution_eq_k}) allows us to determine the areas of existence (continua) of extraordinary waves. In fact, these areas are determined only by the parameter $\EuScript A$ which is substituted in the denominator of Eq.~(\ref{eq:solution_eq_k}). To confirm this, in Fig.~\ref{fig:fig4} we plot the dependence of the parameter $\EuScript A$ on the angles $\theta$ and $\varphi$ since these angles determine the direction of wave propagation in space. The calculations are carried out at two fixed values of $k_0$ which belong to Region~\circled{\small{2}}. The parametric surfaces of $\EuScript A$ are supplemented by plots of corresponding values of the real root $\kappa$ of Eq.~(\ref{eq:disp_eq_k}). At the bottom of each plot, the contours of surfaces $\EuScript A$ and $\kappa$ are drawn, where the continua of extraordinary waves are filled with blue.

From a comparison of data plotted in Figs.~\ref{fig:fig3}(d) and \ref{fig:fig4}(a) as well as in Figs.~\ref{fig:fig3}(e) and \ref{fig:fig4}(b), one can conclude that the number of continua of extraordinary waves and their position on the $\theta-\varphi$ plane is determined by the extrema of the parameter $\EuScript A$. In particular, in Region~\circled{\small{2}} extraordinary waves exist where $\EuScript A<0$, and their continua are centered around the local minima of the parameter~$\EuScript A$.

The condition $\EuScript A=0$ determines the transition between the propagating and non-propagating states of extraordinary waves. From this condition, a set of two equations follows
\begin{equation}
(\varepsilon_{xx}\cos^2\varphi + \varepsilon_{yy}\sin^2\varphi) \sin^2\theta + \varepsilon_{zz} \cos^2\theta = 0,
\label{eq:cond_ep}
\end{equation}
\begin{equation}
(\mu_{xx}\cos^2\varphi + \mu_{yy}\sin^2\varphi) \sin^2\theta + \mu_{zz} \cos^2\theta = 0,
\label{eq:cond_mu}
\end{equation}
which outlines the boundaries of continua of extraordinary waves projected onto the $\theta-\varphi$ plane. For the given frequency and parameters of the superlattice, the equality $\varepsilon_{xx}\approx\varepsilon_{yy}$ holds [see Fig.~\ref{fig:fig2}(a)]. This equality suggests that Eq.~(\ref{eq:cond_ep}) does not depend on $\varphi$. To illustrate this, the solutions of Eqs. (\ref{eq:cond_ep}) and (\ref{eq:cond_mu}) are presented in Fig.~\ref{fig:fig4}(c). One can conclude that the constitutive parameters of the semiconductor subsystem determine the propagation conditions of the wave along a certain polar angle $\theta$, while those of the ferrite subsystem are responsible for the continua variations with the azimuthal angle~$\varphi$.

%%%%%%%%
\subsection{Impact of losses}
\label{sec:losses}

In the vicinity of the frequency of ferromagnetic resonance, the material losses in ferrite become significant and thus should be accounted for. In hyperbolic metamaterials, the presence of losses leads to a change in the form of isofrequency surfaces, which can undergo significant loss-induced modifications.\cite{Ballantine_PhysRevA_2014, Yu_JAP_2016, Fesenko_PhysRevB_2019} Therefore, it is important to study how losses affect dispersion characteristics of bulk waves propagating through the given biaxial gyrotropic medium. As before, our main interest is in revealing the conditions of the extraordinary waves propagation in the frequency range belonging to Region~\circled{\small{2}}.

In the presence of losses, the components of tensors (\ref{eq:eff}) become complex quantities $\varepsilon_{ij}=\varepsilon_{ij}'+ i\varepsilon_{ij}''$ and $\mu_{ij}=\mu_{ij}' + i\mu_{ij}''$ ($i,j=x,y,z$), which yields that the wavevector $\vec{k}$ has complex-valued components $k_j = k_j' + ik_j''$ ($j=x,y,z$). In general, all four roots of Eq.~(\ref{eq:disp_eq_k}) are complex values $\kappa_j= \kappa_j'+ i \kappa_j''$ ($j=1,2,3,4$), and among them, only two solutions are  physically  acceptable. A physically acceptable solution implies that the wave decays as it propagates through a lossy medium. In accordance with the known notations,\cite{Ishimaru_book_1991, Pengchao_NP_2018} two physically acceptable roots of Eq.~(\ref{eq:disp_eq_k}) correspond to the propagation conditions of the `proper' complex modes, while the two remaining roots are related to the `improper' complex modes. In particular, for the chosen time dependence factor $\exp(-i\omega t)$, the wave with $\kappa_j''>0$ satisfies the physically acceptable propagation condition and is related to the proper complex modes, while the wave with $\kappa_j''<0$ has exponentially growing amplitude during propagation and belongs to the improper complex modes. One of the pair of corresponding modes corresponds to the proper or improper extraordinary waves. 

\begin{figure*}[t!]
\centerline{\includegraphics[width=0.9\linewidth]{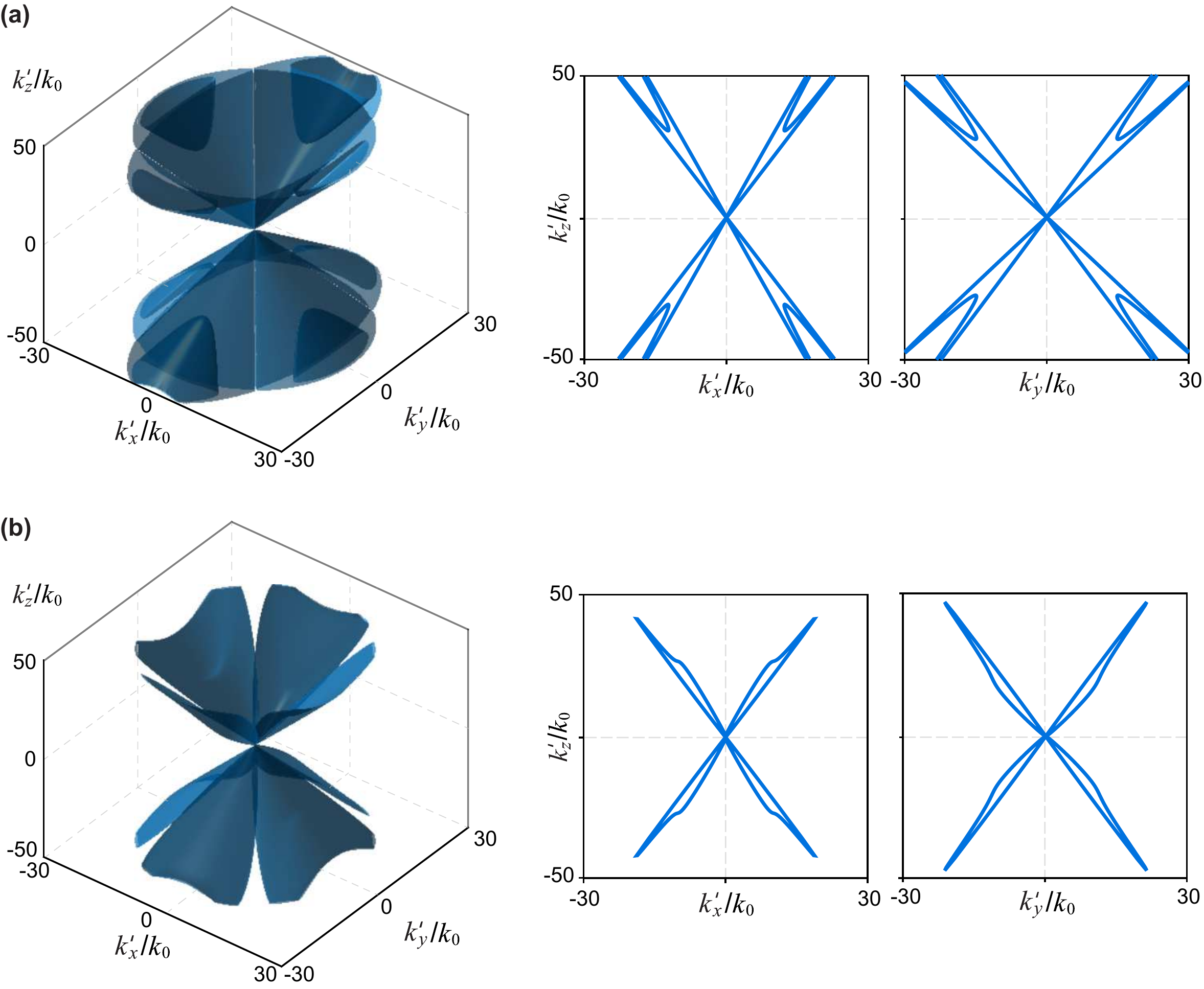}} \caption{Isofrequency surfaces (real parts) and their cross-sections related to propagation conditions of proper extraordinary waves for a lossy biaxial gyrotropic medium. Parameters of losses in the ferrite and semiconductor subsystems are: (a)~$\beta = 1\times 10^{-4}$, $\nu = 2\times10^{-2}$~GHz and (b)~$\beta = 1\times 10^{-3}$, $\nu = 5\times10^{-2}$~GHz. All other parameters of the superlattice are the same as in Fig.~\ref{fig:fig2}. The frequency parameter is fixed at $k_0=149.0$~m$^{-1}$.}
\label{fig:fig5}
\end{figure*}

In our model of the superlattice, the losses are introduced via the parameters $\beta$ and $\nu$ for ferrite and semiconductor layers, respectively (see Appendix~\ref{sec:append}). To illustrate the impact of losses on the propagation conditions of extraordinary waves, the corresponding roots of Eq.~(\ref{eq:disp_eq_k}) are selected, and isofrequency surfaces associated with the real part of $\vec k$ are plotted in Fig.~\ref{fig:fig5} for two different sets of parameters of losses at the fixed frequency. These parameters of losses are typical for actual ferrite and semiconductor materials. Each plot is supplemented by the cross-sections drawn in the $k'_x-k'_z$ and $k'_y-k'_z$ planes. 
% In all plots, we distinguish isofrequency surfaces related to the proper and improper extraordinary waves in different colors.

One can compare the forms of isofrequency surfaces obtained for the lossy medium with that plotted for the lossless case [see Fig.~\ref{fig:fig3}(d)]. From this comparison a conclusion can be made that the hyperbolic topology of the isofrequency surfaces of extraordinary waves survives when the losses are introduced into the system. However, their form appears to be sufficiently distorted. 

Figure~\ref{fig:fig5} suggests that, as soon as a small amount of losses is introduced into  either ferrite or semiconductor subsystem of the superlattice, the isofrequency surface of extraordinary waves no longer extends to infinity, but transits to a closed finite form. This transition is in full compliance with the results reported earlier for the non-magnetic lossy hyperbolic metamaterials.\cite{Ballantine_PhysRevA_2014, Yu_JAP_2016, Gomez_PhysRevLett_2015} As follows from changing in the forms of isofrequency surfaces plotted in Figs.~\ref{fig:fig5}(a) and \ref{fig:fig5}(b), a gradual increase in losses reduces the size of the finite existence areas. Thus, when implementing magnetic control in hyperbolic materials, presence of losses in the constitutive components cannot be ignored, since they strongly affect the forms of isofrequency surfaces for actual structures.

%%%%%%%
\section{Conclusions}
\label{sec:concl}

In conclusion, magnetically induced topological transitions of isofrequency surfaces of bulk waves propagating through an unbounded biaxial gyrotropic medium are studied. The gyrotropic medium is realized from a long-wavelength consideration of a two-component superlattice containing ferrite and semiconductor layers. All our studies were performed in the frequency range near the frequency of ferromagnetic resonance. 

In the lossless case, the topological transitions of isofrequency surfaces of extraordinary high-$k$ waves from an open type-I hyperboloid to the form of a cone cut into either two or four parts as well as a bi-hyperboloid are demonstrated. Such intricate forms of topology are possible when principal components of both permittivity and permeability tensors become negative quantities. For such an extremely anisotropic medium, the areas of existence (continua) of extraordinary waves are defined, and the conditions which determine the transition between the propagating and non-propagating states of these waves are revealed. The forms of isofrequency surfaces obtained for the lossless case broaden our knowledge on the possible topologies of dispersion existing in anisotropic materials.

The presence of losses leads to a strong distortion of hyperbolic isofrequency surfaces as compared with the lossless case. Therefore, when gaining magnetic control in hyperbolic metamaterials, in addition to realizing the desired topology of isofrequency surfaces, it is also necessary to analyze obtained solution on its physical feasibility, as soon as a system with realistic losses is under consideration.

%%%%%%
\appendix
\section{Constitutive parameters of ferrite and semiconductor layers}
\label{sec:append}

The expressions for tensors components of the underlying constitutive parameters of magnetic $\hat \mu_m\to\hat g^{(m)}$ and semiconductor $\hat \varepsilon_s\to\hat g^{(s)}$ layers can be written in the form:
\begin{equation}
\hat g^{(j)}=\left( {\begin{matrix}
   {g_1} & {ig_2} & {0} \cr
   {-ig_2} & {g_1} & {0} \cr
   {0} & {0} & {g_3} \cr
\end{matrix}
} \right). 
\label{eq:gfs}
\end{equation}

For magnetic layers\cite{Gurevich_book_1963, Collin_book_1992} the components of tensor $\hat g^{(m)}$ are: $g_1=1+\chi' + i\chi''$, $\quad g_2=\Omega'+i\Omega''$, $ g_3=1$, and $\quad\chi'=\omega_0\omega_m[\omega^2_0-\omega^2(1-\beta^2)]D^{-1}$, $\chi''=\omega\omega_m \beta[\omega^2_0+\omega^2(1+\beta^2)]D^{-1}$, $\quad\Omega'=\omega\omega_m[\omega^2_0-\omega^2(1+\beta^2)]D^{-1}$, $\Omega''=2\omega^2\omega_0\omega_m \beta D^{-1}$, $\quad D=[\omega^2_0-\omega^2(1+\beta^2)]^2+4\omega^2_0\omega^2 \beta^2$, where $\omega_0$ is the Larmor frequency and $\beta$ is the dimensionless damping constant.

For semiconductor layers\cite{akhiezer1975plasma} the components of tensor $\hat g^{(s)}$ are: $g_1=\varepsilon_l[ {1-\omega_p^2 (\omega+i\nu)[\omega((\omega+i\nu)^2-\omega_c^2)]^{-1}}]$, $g_2=\varepsilon_l\omega_p^2\omega_c[\omega((\omega+i\nu)^2-\omega_c^2)]^{-1}$, $g_3=\varepsilon_l[1-\omega_p^2[\omega(\omega+i\nu)]^{-1}]$, where $\varepsilon_l$ is the part of permittivity of the lattice, $\omega_p$ is the plasma frequency, $\omega_c$ is the cyclotron frequency and $\nu$ is the electron collision frequency in plasma.

Relative permittivity $\varepsilon_m$ of the ferrite layers as well as relative permeability $\mu_s$ of the semiconductor layers are scalar quantities.

\bibliography{tetra}

\end{document}